\documentstyle[preprint,aps]{revtex}
\tightenlines
\begin{document}
\draft
\title
{Duality and anyonic excitations}
\author
{Wellington da Cruz\footnote{E-mail: wdacruz@fisica.uel.br}} 
\address
{Departamento de F\'{\i}sica,\\
 Universidade Estadual de Londrina, Caixa Postal 6001,\\
Cep 86051-970 Londrina, PR, Brazil\\}
\date{\today}
\maketitle
\begin{abstract}
We consider anyonic excitations classified into 
equivalence classes labeled by Hausdorff dimension, $h$ and 
introduce the concept of duality between such classes, 
defined by $\tilde{h}=3-h$. In this way, we confirm that 
the filling factors for which 
the Fractional Quantum Hall Effect ( FQHE ) were 
observed just appears into these classes and 
the duality in case is between quasihole and 
quasiparticle excitations for these FQHE systems. Exchanges of dual 
pairs $\left(\nu,\tilde{\nu}\right)$, suggests conformal invariance.  
\end{abstract}


We have classified the fractional spin particles or 
anyonic excitations into equivalence 
classes labeled by Hausdorff dimension, $h$ and 
conjectured in this way a new hierarchy scheme 
for the filling factors, $\nu$. Such parameters characterize 
the Fractional Quantum Hall Effect ( FQHE ) and second 
our approach we can predicting for which values of $\nu$ 
FQHE can be observed\cite{R1}. In \cite{R2} we have obtained 
an anyonic distribution function for each class $h$, which 
reduces to bosonic and fermionic distributions, when $h=2$ and 
$h=1$, respectively.

Now, we introduce the concept of duality between 
equivalence classes, defined by 

\begin{equation} 
\label{e.1}
\tilde{h}=3-h,
\end{equation}

\noindent such that, for $h=1$, we have $\tilde{h}=2$ and 
for $h=2$, we have $\tilde{h}=1$. This means that for $h$ 
defined into the interval $1< h < 2$, with $h$ a rational 
fraction with odd denominator, is 
related to the filling factor $\nu$ as follows:
 
\begin{eqnarray}
\label{e.2}
&&h_{1}=2-\nu,\;\;\;\; 0 < \nu < 1;\;\;\;\;\;\;\;\;
 h_{2}=\nu,\;\;\;\;
\;\;\;\;\;\;\;\;\; 1 <\nu < 2;\;\nonumber\\
&&h_{3}=4-\nu,\;\;\;\; 2 < \nu < 3;\;\;\;\;\;\;\;\;
h_{4}=\nu-2,\;\;\;\;\;\;\; 3 < \nu < 4;\;\\
&&h_{5}=6-\nu,\;\;\;\; 4 < \nu < 5;\;\;\;\;\;\;\;
h_{6}=\nu-4,\;\;\;\;\;\;\;\; 5 < \nu < 6;\nonumber\\
&&etc.\nonumber
\end{eqnarray}

\noindent For the set of values of the filling factors 
$\nu$ experimentally observed\cite{R3}, second 
our relations ( Eq.\ref{e.2} and Eq.\ref{e.1} ), we get the 
classes, $h$ and $\tilde{h}$:

\begin{eqnarray}
&&\left\{\frac{1}{3},\frac{5}{3},\frac{7}{3},
 \frac{11}{3},\cdots\right\}_
{h=\frac{5}{3}},\;\;\;\;\;\;\;\;\;\;\left\{\frac{2}{3},
\frac{4}{3},\frac{8}{3},\frac{10}{3},\cdots\right\}_
{{\tilde{h}}=\frac{4}{3}};\nonumber\\
&&\left\{\frac{1}{5},\frac{9}{5},\frac{11}{5},
 \frac{19}{5},\cdots\right\}_
{h=\frac{9}{5}},\;\;\;\;\;\;\;\;\left\{\frac{4}{5},
\frac{6}{5},\frac{14}{5},\frac{16}{5},\cdots\right\}_
{{\tilde{h}}=\frac{6}{5}};\nonumber\\
&&\left\{\frac{2}{7},\frac{12}{7},\frac{16}{7},
 \frac{26}{7},\cdots\right\}_
{h=\frac{12}{7}},\;\;\;\;\;\left\{\frac{5}{7},
\frac{9}{7},\frac{19}{7},\frac{23}{7},\cdots\right\}_
{{\tilde{h}}=\frac{9}{7}};\nonumber\\
&&\left\{\frac{2}{9},\frac{16}{9},\frac{20}{9},
 \frac{34}{9},\cdots\right\}_
{h=\frac{16}{9}},\;\;\;\;\;\left\{\frac{7}{9},
\frac{11}{9},\frac{25}{9},\frac{29}{9},\cdots\right\}_
{{\tilde{h}}=\frac{11}{9}};\nonumber\\
&&\left\{\frac{2}{5},\frac{8}{5},\frac{12}{5},
 \frac{18}{5},\cdots\right\}_
{h=\frac{8}{5}},\;\;\;\;\;\;\;\;\left\{\frac{3}{5},
\frac{7}{5},\frac{13}{5},\frac{17}{5},\cdots\right\}_
{{\tilde{h}}=\frac{7}{5}};\\
&&\left\{\frac{3}{7},\frac{11}{7},\frac{17}{7},
 \frac{25}{7},\cdots\right\}_
{h=\frac{11}{7}},\;\;\;\;\;\left\{\frac{4}{7},
\frac{10}{7},\frac{18}{7},\frac{24}{7},\cdots\right\}_
{{\tilde{h}}=\frac{10}{7}};\nonumber\\
&&\left\{\frac{4}{9},\frac{14}{9},\frac{22}{9},
 \frac{32}{9},\cdots\right\}_
{h=\frac{14}{9}},\;\;\;\;\;\left\{\frac{5}{9},
\frac{13}{9},\frac{23}{9},\frac{31}{9},\cdots\right\}_
{{\tilde{h}}=\frac{13}{9}};\nonumber\\
&&\left\{\frac{6}{13},\frac{20}{13},\frac{32}{13},
 \frac{46}{13},\cdots\right\}_
{h=\frac{20}{13}},\;\;\;\left\{\frac{7}{13},
\frac{19}{13},\frac{33}{13},\frac{45}{13},\cdots\right\}_
{{\tilde{h}}=\frac{19}{13}};\nonumber\\
&&\left\{\frac{5}{11},\frac{17}{11},\frac{27}{11},
 \frac{39}{11},\cdots\right\}_
{h=\frac{17}{11}},\;\;\;\left\{\frac{6}{11},
\frac{16}{11},\frac{28}{11},\frac{38}{11},\cdots\right\}_
{{\tilde{h}}=\frac{16}{11}};\nonumber\\
&&\left\{\frac{7}{15},\frac{23}{15},\frac{37}{15},
 \frac{53}{15},\cdots\right\}_
{h=\frac{23}{15}},\;\;\;\left\{\frac{8}{15},
\frac{22}{15},\frac{38}{15},\frac{52}{15},\cdots\right\}_
{{\tilde{h}}=\frac{22}{15}}.\nonumber
\end{eqnarray}

\noindent We note that in each class, the first 
filling factors are the experimental values observed ( also some 
second and third values ) 
that is, the Hall resistance  develops plateaus in these 
quantized values, which are related to the fraction of 
electrons that form collective excitations as 
quasiholes or quasiparticles in FQHE systems. The relation 
of duality between equivalence classes labeled by $h$, 
therefore, can indicate a way as determine the dual 
of a specific value of $\nu$ ( or ${\tilde{\nu}}$ ) observed. 
On the other hand, for excitations above the 
Laughlin ground state, the exchange of two 
quasiholes\cite{R4} with coordinates $z_{\alpha}$ and 
$z_{\beta}$ produces the condition on the phase

\begin{equation}
\label{e.3}
\exp\left\{\imath\pi \nu_{1}\right\}=
\exp\left\{\imath\pi\frac{1}{m}\right\},
\end{equation}
\noindent with $\nu_{1}$$=$$\frac{1}{m}+2p_{1}$; 
and for a second generation 
of quasihole excitations, the effective wavefunction 
carries the factor $\left(z_{\alpha}-z_{\beta}\right)^{\nu_{2}}$, 
with $\nu_{2}$$=$$\frac{1}{\nu_{1}}+2p_{2}$; $m=3,5,7,\cdots$ 
and $p_{1}$,$\;$$p_{2} $ are positive integers. 
In \cite{R5} we have noted that these conditions 
over the filling factor $\nu$ confirm our 
classification of the collective excitations in terms of $h$. Another 
interesting point is that in each class, we have more filling factors 
which those generate by ( Eq.\ref{e.3} ), that is, our classification 
cover the complete spectrum of states. Now, 
we can see that the duality between equivalence classes 
means duality between quasiholes and quasiparticles, 
that is, between $h$ and $\tilde{h}$. Therefore, as we 
said elsewhere\cite{R1}, $h$ tell us about the 
nature of the anyonic excitations.

On the other hand, we note that the anyonic exchanges of duals get 
a phase difference, modulo constant,

\begin{equation}
\left|\nu-\tilde{\nu}\right|=\left|\Delta\nu\right|=h-\tilde{h}=const,
\end{equation}

\noindent suggesting an invariance, conformal symmetry. We have also for 
the elements of $h$ and $\tilde{h}$, the following relations

\begin{eqnarray}
\label{e.9}
&&\frac{\nu_{i+1}-\nu_{i}}{\tilde{\nu}_{j+1}-\tilde{\nu}_{j}}=1;\\
&&\frac{\nu_{j+1}-\nu_{j}}{\tilde{\nu}_{i+1}-\tilde{\nu}_{i}}=1,\nonumber
\end{eqnarray}

\noindent with $i=1,3,5,etc.$ and $j=2,4,6,etc.$; such that the pairs 
$\left(i,j\right)=\left(1,2\right),\left(3,4\right),etc.$ satisfy the 
expressions ( Eq.\ref{e.9} ).


\begin{thebibliography}{99}
\bibitem{R1} W. da Cruz, preprint/UEL-DF/W-03/97, 
hep-th/9802135; ibid, preprint/UEL-DF/W-04/97, 
cond-mat/9802266.
\bibitem{R2} W. da Cruz, preprint/UEL-DF/W-01/98, 
hep-th/9802123.
\bibitem{R3} A. H. MacDonald, cond-mat/9410047 and
references therein; A. M. Chang in {\it The Quantum Hall Effect} 
( Springer Verlag, 1990 ), Ed. by R. E. Prange and S. M. Girvin
and references therein; {\it Perspectives in Quantum Hall Effects}, Ed. by 
Sankar Das Sarma and Aron Pinczuk ( Wiley, New York, 1997 ); 
T. Chakraborty and P. Pieti\"ainen, {\it The Fractional 
Quantum Hall Effect: Perspectives of an 
incompressible quantum fluid} ( Springer-Verlag, New York, 1988 ), 
Springer Series in Solid State Sciences, {\bf 85}; {\it 
Quantum Hall Effect: A Perspective }, Ed. by A. H. 
MacDonald ( Kluwer, Boston, 1989 ).   
\bibitem{R4} A. Lerda, {\it Anyons}, Lectures Notes in Physics 
( Springer Verlag, 1992 ); R. B. Laughlin, Phys. Rev. Lett. 
{\bf 50} 1395 ( 1983 ); ibid, Phys. Rev. {\bf B23}, 3383 
(1983).
\bibitem{R5} W. da Cruz, preprint/UEL-DF/W-05/97, cond-mat/9802267. 

\end{thebibliography}
\end{document}